\documentclass[aps,prd,nofootinbib,twocolumn,superscriptaddress,letterpaper,preprintnumbers]{revtex4-1}

\usepackage{amsmath, amssymb,braket}
\usepackage[dvipsnames]{xcolor}
\usepackage{color}
\usepackage{float}
\usepackage{graphicx}
\usepackage{subfigure}
\usepackage{natbib}
\usepackage{adjustbox}
\usepackage[colorlinks=true
,urlcolor=blue
,anchorcolor=blue
,citecolor=blue
,filecolor=blue
,linkcolor=red
,menucolor=blue
,hyperfootnotes=false,
,linktocpage=true
,pdfproducer=medialab
,pdfa=true
]{hyperref}

\newcommand{\D}{{\rm d}}
\newcommand\orcid[1]{\href{http://orcid.org/#1}{\adjustbox{trim={-.15\width} {0\height} {-.15\width} {0\height},clip}{\includegraphics[height=10pt]{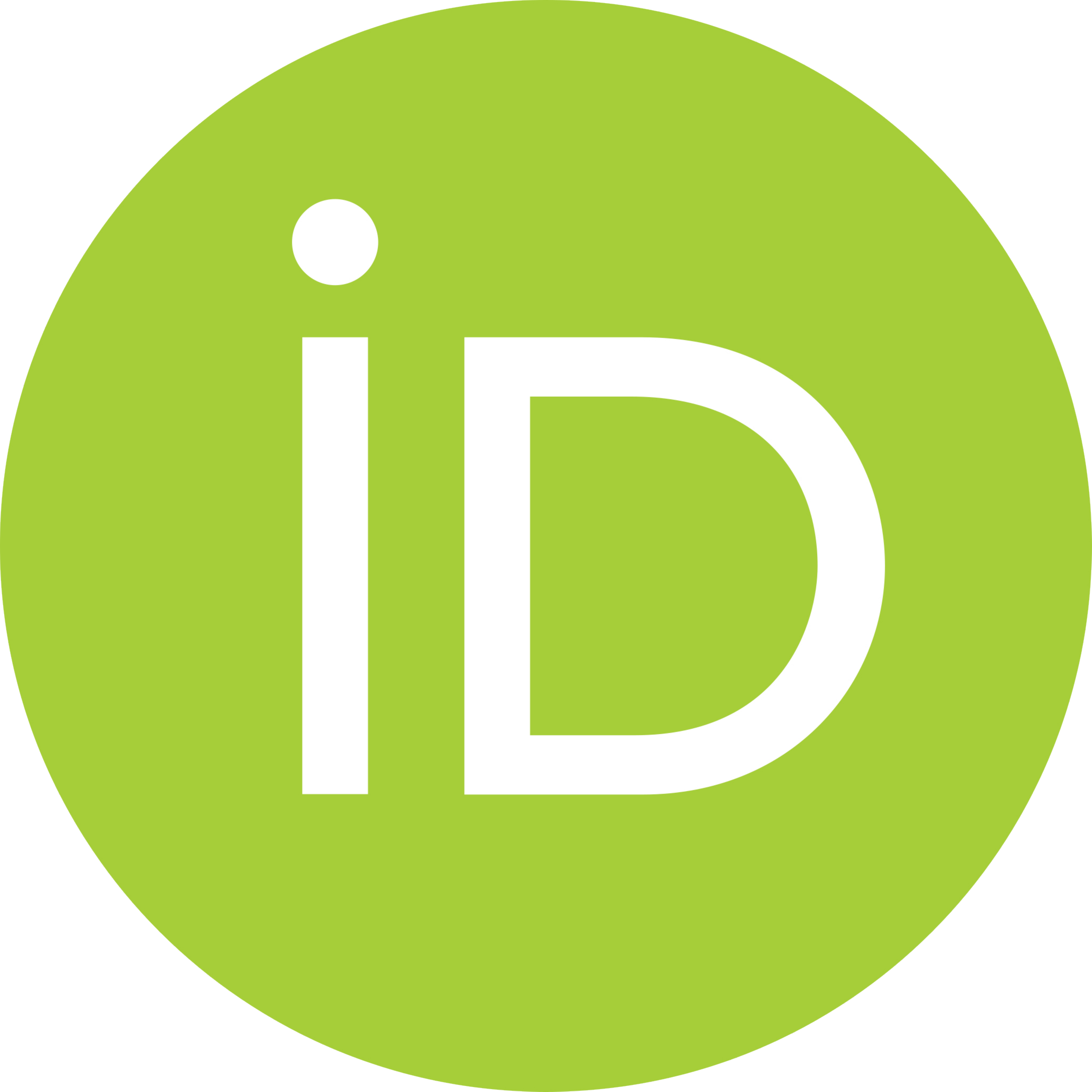}}}}

\begin{document}

\title{Not-quite-primordial black holes}

\author{Wenzer Qin\orcid{0000-0001-7849-6585}}
\affiliation{Center for Cosmology and Particle Physics, Department of Physics, New York University, New York, NY 10003, USA}

\author{Soubhik Kumar\orcid{0000-0001-6924-3375}}
\affiliation{Institute of Cosmology, Department of Physics and Astronomy, Tufts University, Medford, MA
02155, USA}
\affiliation{Center for Cosmology and Particle Physics, Department of Physics, New York University, New York, NY 10003, USA}

\author{Priyamvada Natarajan\orcid{0000-0002-5554-8896}} 
\affiliation{Department of Astronomy, Yale University, 219 Prospect Street, New Haven, CT 06511, USA}
\affiliation{Department of Physics, Yale University, New Haven, CT 06520, USA}
\affiliation{Yale Center for the Invisible Universe, New Haven, CT 06511, USA}
\affiliation{Black Hole Initiative, Harvard University, 20 Garden Street, Cambridge, MA 02138, USA}

\author{Neal Weiner\orcid{0000-0003-2122-6511}} 
\affiliation{Center for Cosmology and Particle Physics, Department of Physics, New York University, New York, NY 10003, USA}
\affiliation{Center for Computational Astrophysics, 160 5th Avenue, New York, NY 10010, USA}
\affiliation{Theoretical Physics Department, CERN, 1211 Geneva, Switzerland}

\begin{abstract}
    We propose a new mechanism for the formation of seeds of supermassive black holes at early cosmic epochs. 
    Our scenario explores density fluctuations that are enhanced relative to $\Lambda$CDM expectations, but with amplitudes that are not large enough to form primordial black holes, and that can still lead to collapsed dark matter halos at very early times.
    For halos forming prior to $1+z \approx 200$, the Cosmic Microwave Background (CMB) is energetic enough to suppress the formation of molecular hydrogen, hence preventing cooling and fragmentation, as a consequence of which baryons falling into the potential well of the halo may undergo ``direct collapse" into a black hole. We show using a few illustrative models how this mechanism may account for the abundance of high-redshift black holes inferred from observations by the \textit{James Webb Space Telescope} while remaining consistent with limits from CMB spectral distortions. Limits on the primordial power spectrum are also derived by requiring that the universe not re-ionize too early.
\end{abstract}

\maketitle

\textbf{Introduction}---
The standard cosmological model posits gravity-driven growth of nearly scale-invariant perturbations that collapse upon reaching nonlinearity. This model for structure formation has been successfully tested empirically on a wide range of scales directly down to halo masses of $M_h\sim 10^9 M_\odot$~\cite{Bechtol:2022koa}. Below this mass scale, there are at present fewer observational constraints.

There are reasons, both theoretical and observational, for interest in probing smaller mass scales even below $10^5M_{\odot}$. Many models of inflation and dark matter can produce density enhancements at smaller scales, which may have important consequences at later cosmic times. These lower mass scales are of great interest as they could provide yet another channel for the formation of black hole (BH) seeds that grow into present-day supermassive black holes (SMBHs). In particular, recent \textit{James Webb Space Telescope} (JWST) observations of ``little red dots'' (LRDs)~\cite{2023ApJ...954L...4K,2023Natur.616..266L,2023ApJ...959...39H,2024A&A...691A.145M,2024ApJ...963..129M}, accreting black holes at $z>10$ ~\cite{2024Natur.627...59M,2024NatAs...8..126B,2024ApJ...960L...1N}, and galaxy candidates at $1+z \geq 17$~\cite{2025arXiv250315594P} may indicate that the formation of BH seeds could be more expansive and might occur earlier than expected in conventional structure formation models~\cite{2011BASI...39..145N,2024A&A...690A.182D,Zhang:2025asq,Matteri:2025vnv}.

In this {\it letter}, we propose that large density enhancements on very small scales could produce an early population of BH seeds, providing a new formation channel at very high redshifts. Over-densities collapsing at $1+z \gtrsim 200$, when the CMB suppresses molecular cooling, are expected to cool only via atomic hydrogen. 
This inhibits fragmentation and favors direct collapse into BHs, akin to proposals made for later cosmic epochs at $z \sim 20-30$ ~\cite{2006MNRAS.371.1813L}. 
In the following sections, we review the conditions for forming direct collapse black holes (DCBHs) and consider the conditions under which large overdensities might collapse at very high redshifts, leading to the formation of a population of SMBH seeds.

\begin{figure}
    \centering
    \includegraphics[width=\columnwidth]{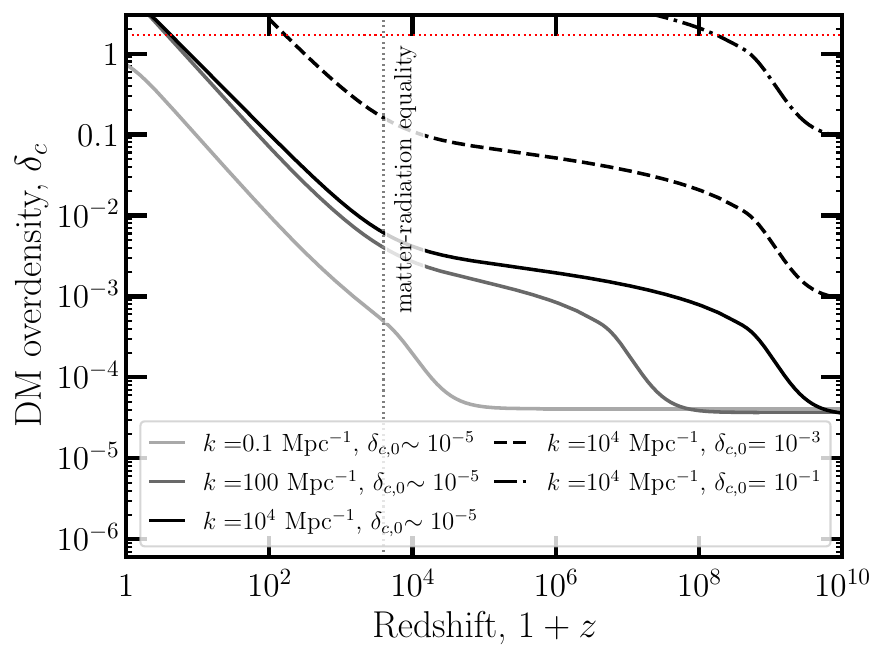}
    \caption{
    Linear evolution of density perturbations. When varying the wavenumber at which a given mode of fluctuations re-enters the horizon (solid lines), the redshifts at which the overdensities enter the nonlinear regime all occur at $1+z < 10$. In contrast, varying the initial amplitude of the overdensity (black lines), one can achieve proportional changes to the redshift of onset of nonlinearity, so long as nonlinear evolution begins in the matter-dominated era. These curves were produced using \texttt{CLASS}~\cite{2011arXiv1104.2932L,2011JCAP...07..034B}.
    }
    \label{fig:delta_evolution}
\end{figure}

\textbf{Direct collapse black holes}---In standard $\Lambda$CDM cosmology, the initial conditions for density fluctuations, $\delta = (\rho - \overline{\rho}) / \overline{\rho}$, are characterized by a nearly scale-invariant power spectrum seeded by, e.g., single-field slow-roll inflation~\cite{Achucarro:2022qrl}. These initial fluctuations are quite small, given that the amplitude of the power spectrum has been measured to be $A_s \sim 2 \times 10^{-9}$~\cite{Planck:2018vyg}, and hence the typical amplitude of an initial fluctuation is a ${\rm few} \times 10^{-5}$. 
In the spherical collapse model, an overdensity collapses once linear evolution predicts $\delta \sim 1.69$, meaning only modes that entered the horizon during radiation domination can grow enough to become non-linear by today.
Fig.~\ref{fig:delta_evolution} shows in solid lines the linear evolution of typical overdensities with a few different wavenumbers. Despite spanning five orders of magnitude in $k$, all modes collapse at similar redshifts, $1+z \lesssim 10$, due to their slow growth during radiation domination.

Therefore, according to this standard picture, the first generation of compact objects is expected to begin to form after $1+z \sim 20-30$ as demonstrated in cosmological simulations of early structure formation \cite{Bromm:2003vv, Bromm:2013iya, Klessen:2023qmc}. Distinguishing between dark matter, $\delta_c$, and baryonic components of the overdensity, we assume that dark matter forms a virialized halo shortly after $\delta_c$ grows to be greater than one and provides a gravitational potential well for the subsequent collapse of the baryons. 
Once the baryons also virialize, their temperature is given by~\cite{2010gfe..book.....M}
\begin{equation}
    T_\mathrm{vir} \approx 9700 \,\mathrm{K} \, 
    \left( \frac{M_h}{10^7 M_\odot} \right)^{2/3} 
    \left( \frac{\rho}{4 \times 10^{-24} \,\mathrm{g}\,\mathrm{cm}^{-3}} \right)^{1/3}
    \label{eqn:Tvir}
\end{equation}
where we assume that the gas has a mean molecular weight of $\mu = 0.6$ and $\rho \sim 4 \times 10^{-24} \,\mathrm{g}\,\mathrm{cm}^{-3}$ is the density of a halo virializing at $1+z \sim 20$. 
However, baryons, unlike dark matter, can cool, and thus can collapse well beyond the virial density.

For the first generation of compact objects, the relevant cooling processes are radiative cooling by atomic and molecular hydrogen~\cite{Barkana:2000fd}. Neutral atomic hydrogen is abundant after recombination and cools efficiently above temperatures of $10^4$ K~\cite{Barkana:2000fd,Oh:2001ex}. While molecular hydrogen enables more efficient cooling to lower temperatures, its global abundance is suppressed until $z \sim 120$, when the CMB cools enough that photodetachment of the intermediate H$^-$ becomes inefficient and allows an appreciable amount of H$_2$ to form~\cite{Hirata:2006bt}---the inclusion of expected nonthermal photons further suppresses the cosmic abundance of H$_2$~\cite{2013MNRAS.434..114C} during the dark ages.

To contrast how these two cooling mechanisms impact the collapse of baryons, we use the Jeans mass, which serves as a heuristic for the typical mass of collapsed objects. We derive the Jeans mass in terms of the gas temperature and density by inverting Eq.~\eqref{eqn:Tvir}, 
\begin{equation}
    M_J = 2 \times 10^7 \,M_\odot \, \left( \frac{T}{10^4 \,\mathrm{K}} \right)^{3/2} \left( \frac{\rho}{4 \times 10^{-24} \,\mathrm{g}\,\mathrm{cm}^{-3}} \right)^{-1/2}
    \label{eqn:MJ}.
\end{equation}
For the redshifts of interest, which we will find in the following to be $z\gtrsim 200$, the global gas temperature follows the CMB temperature, so $M_J$ is approximately constant with the redshift.
For a halo virializing at $T \geq 10^4$ K, atomic cooling will keep the temperature at that threshold. 
If H$_2$ is sufficiently abundant, then molecular cooling will further lower the temperature to $\mathcal{O} (100 \,\mathrm{K})$, causing the Jeans mass to drop precipitously and the gas to fragment, which is expected to catalyze star formation.

However, if molecular cooling is suppressed, the gas will condense isothermally at $10^4$ K and the Jeans mass only falls as $M \sim \rho^{-1/2}$, with most of the gas collapsing into a monolithic central object that will likely lead to the formation of a black hole of mass $\sim 10^5 \,M_\odot$~\cite{Eisenstein:1994nh}. Defining the molecular hydrogen fraction as $x_{\mathrm{H}_2} = n_{\mathrm{H}_2} / n_\mathrm{H}$, where $n_{\mathrm{H}_2}$ is the number density of molecular hydrogen and $n_{\mathrm{H}}$ the total number density of hydrogen nuclei, we show in the end matter that halos at these redshifts with $T_\mathrm{vir} > 10^4$~K must have $x_{\mathrm{H}_2} \lesssim 10^{-7}$ in order for molecular cooling to be inefficient, and therefore only halos collapsing at $1+z > 200$ are capable of ``direct collapse".

Angular momentum also plays a critical role, as high-spin halos may not overcome the angular momentum barrier and thus fail to collapse~\cite{Bromm:2002hb,Koushiappas:2003zn,2006MNRAS.371.1813L}.
Hence, the key conditions that are required to effect direct collapse are limited cooling and low spin, which allow the gas to collapse without significant fragmentation, giving rise to relatively massive seed BHs. 
Unlike the high-redshift scenario ($z \sim 200$) considered here, at $z \sim 20$–30, 
direct collapse can also be realized through suppression of molecular cooling by external Lyman-Werner radiation~\cite{2006MNRAS.371.1813L,2012MNRAS.425.2854A}, dynamical heating~\cite{2019Natur.566...85W}, and turbulence from cold flows~\cite{Latif:2022vwc} or mergers~\cite{2010Natur.466.1082M,2024ApJ...961...76M}.

\textbf{Primordial black holes}---Primordial black holes (PBHs) forming soon after inflation could also serve as putative seeds for SMBHs~\cite{1974MNRAS.168..399C}. PBHs form from initial density fluctuations that are so large in amplitude that they almost immediately collapse into BHs upon horizon reentry~\cite{Carr:1975qj}. During the radiation-dominated era, this requires $\delta > c_s^2 = 1/3$, where $c_s$ is the sound speed. Within $\Lambda$CDM, such fluctuations are exponentially rare, so producing a significant abundance of PBHs requires enhancing density fluctuations by several orders of magnitude, which demands fine-tuning of almost any model~\cite{Carr:2020xqk,Green:2020jor, Cole:2023wyx}. 
Moreover, such large fluctuations lead to other observational consequences, such as CMB spectral distortions~\cite{Chluba:2012we,Kohri:2014lza,Nakama:2017xvq,Unal:2020mts,Sharma:2024img,Byrnes:2024vjt,Pritchard:2025yda}. Enhancing density fluctuations above the critical threshold for PBHs implies that there also exist many fluctuations that are enhanced relative to $\Lambda$CDM expectations but that do not meet this stringent threshold. The evolution of such intermediate fluctuations has interesting downstream consequences that we discuss next. Previous works have studied related scenarios resulting in smaller ultra-dense minihalos~\cite{Ricotti:2009bs,2023MNRAS.520.4370D} or late-forming PBHs from long-range forces~\cite{Chakraborty:2022mwu,Flores:2020drq}.

\begin{figure}
    \centering
    \includegraphics[width=\columnwidth]{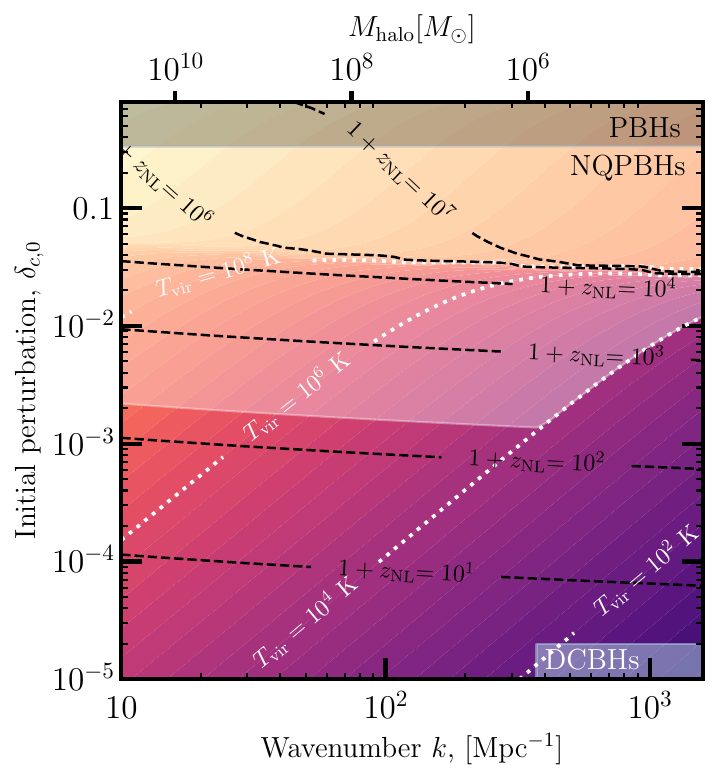}
    \caption{
    The temperature and redshift of a halo at virialization as a function of wavenumber and initial perturbation. The pale white region, bounded from below by $1+z_{\rm NL}=200$ and $T_{\rm vir}=10^4$~K contours, shows where it may be possible to produce ``not-quite-primordial black holes".
    }
    \label{fig:summary}
\end{figure}

\textbf{Not-quite-primordial}---Consider a dark matter overdensity with an initial value of $\delta_{c,i} \sim 10^{-3}$, which is significantly lower than the requirement for PBH formation. Compared to a typical overdensity in $\Lambda$CDM ($\delta_{c,i} \sim {\rm few} \times 10^{-5}$) with the same wavenumber, this enhanced overdensity will receive the usual amount of growth through radiation domination as without the enhancement. Hence, the change in the redshift at which the overdensity virializes is determined by its linear growth with the scale factor, $\delta_c \sim a$, during matter domination. If a $\Lambda$CDM fluctuation virializes at $1+z_\mathrm{vir}$, then the enhanced fluctuation will virialize earlier, at $10^{-3}/(\text{few} \times 10^{-5}) \times (1+z_\mathrm{vir})$. 
In other words, a halo forming at $1+z \sim 20$ in $\Lambda$CDM would instead collapse at $1+z \sim 200$ if $\delta_{c,i}$ is increased by a factor of 10, showing even modest enhancements can trigger earlier collapse.

We illustrate this idea in Fig.~\ref{fig:delta_evolution}, which shows the linear evolution of three density perturbations with different initial values at $k = 10^4$ Mpc$^{-1}$. The perturbation with $\delta_{c,i} = \rm{few} \times 10^{-5}$ evolves into the non-linear regime around $1+z = \mathcal{O} (1)$, while this occurs for the perturbation with $\delta_{c,i} = 10^{-3}$ closer to $1+z \sim 100$. For a large enough initial value $\delta_{c,i} = 10^{-1}$ (as is the case for PBHs), the perturbation instead evolves into the nonlinear regime during the radiation dominated era itself, shortly after entering the horizon.

If the dark matter halo forms prior to $1+z \sim 200$, before H$_2$ is abundant, then the baryons in such halos may only undergo atomic cooling (see the end matter for a comparison of this redshift to Refs.~\cite{2015ApJ...814...18H,2024PASJ...76..850I}). 
As discussed previously, if the gas also has low angular momentum, then the halo meets the necessary conditions to form a DCBH. Finally, baryons can only fall into the gravitational potentials of dark matter if their relative streaming velocity is small enough, which is true for most halos after $1+z \sim 400$; see the end matter for details. 
We refer to black holes forming by this mechanism as ``not-quite-primordial black holes" (NQPBHs), to emphasize that these objects arise from enhanced initial fluctuations---similar to PBHs---but the required enhancement is smaller. 
Thus, NQPBHs form later than PBHs but prior to the DCBHs in the standard picture, and the physics driving their formation is direct collapse.

In summary, the formation of an NQPBH in an early dark matter halo requires three concurrent conditions:
\begin{itemize}
    \item The dark matter density should be large enough for the halo to form at $1+z \gtrsim 200$, so that the H$_2$ abundance is suppressed. Although we consider adiabatic perturbations in this work, this condition could also be met with dark matter isocurvature perturbations---we leave a detailed exploration to future work.

    \item The halo mass be large enough for atomic cooling to be efficient, e.g., $T_\mathrm{vir} \gtrsim 10^4$ K.

    \item The halo spin be low enough for mass to concentrate rapidly at the center of the halo.
\end{itemize}
Fig.~\ref{fig:summary} shows the density fluctuations for which the first two conditions are viable.  Given the initial amplitude and wavenumber, we show contours for the redshift at which the linearly evolving overdensity exceeds unity, and the virial temperature. 
On the top axis, we also show the halo mass, corresponding to the mass contained within the horizon (e.g. within a radius $2\pi/k$) when the mode with wavenumber $k$ re-enters, which sets the \textit{maximum} initial mass of the NQPBH. For $k\lesssim 400$ Mpc$^{-1}$, the mass contained within the halo is large enough that all halos forming before $1+z \sim 200$ have a virial temperature higher than $10^4$ K, hence NQPBHs may form for $\delta_{c,i} \gtrsim 10^{-3}$. For larger wavenumbers, the stronger condition is $T_\mathrm{vir} > 10^4$ K and $\delta_{c,i}$ must be larger to form NQPBHs, reaching $\delta_{c,i} > 10^{-2}$ for $k = 10^3$ Mpc$^{-1}$.

\textbf{Illustrative models}---The formation of NQPBHs depends on having sufficiently enhanced matter overdensities with low angular momentum; here we show how this can be realized in the context of a few different models. For each model, we show that one can achieve BH seed abundances that account for the number densities of SMBHs inferred from JWST observations at later times, while remaining consistent with current constraints.

To calculate the number density of NQPBHs, we look at the distribution of halos that have formed by $1+z = 200$ across mass and angular momentum, $\D^2 n /(\D M \,\D\lambda) $, where $\lambda$ is the dimensionless spin parameter defined in the standard way, $\lambda = J |E|^{1/2}/(G M^{5/2})$, and then evolve the BH mass to $1+z = 8$ by assuming the central black hole is an order-unity fraction of the halo mass and fixing its accretion history. We then integrate the angular momentum and mass so that the halo is large enough to both virialize at $T > 10^4$ K and yield a BH with a luminosity high enough to be observed by JWST:
\begin{align}
    n_\mathrm{BH} &= \int_{M_\mathrm{min}}^\infty \D M \, \int_0^{\lambda_\mathrm{max}} \D\lambda ~\kappa~\frac{\D^2 n}{\D M \,\D\lambda}\bigg \rvert_{1+z = 200}.
    \label{eqn:nBH}
\end{align}
To determine $M_\mathrm{min}$ and $\kappa\equiv \D M_{1+z = 200} / \D M_{1+z = 8}$, we assume the standard Eddington-limited accretion, which is described in the end matter. We integrate the spin parameter up to a cutoff value, $\lambda_\mathrm{max}$. While the spin distribution of very high redshift halos is not known, we consider two different distributions ~\cite{2010MNRAS.407..691D} to bracket this uncertainty, which are also described in the end matter. 
Then, given an initial matter power spectrum, we can determine the initial black hole mass function $\D n/\D M$ using the extended Press-Schechter formalism~\cite{1974ApJ...187..425P,1991ApJ...379..440B}. In what follows, we explore the results of assuming different forms for the power spectrum.

\begin{figure}
    \centering
    \includegraphics[width=\columnwidth]{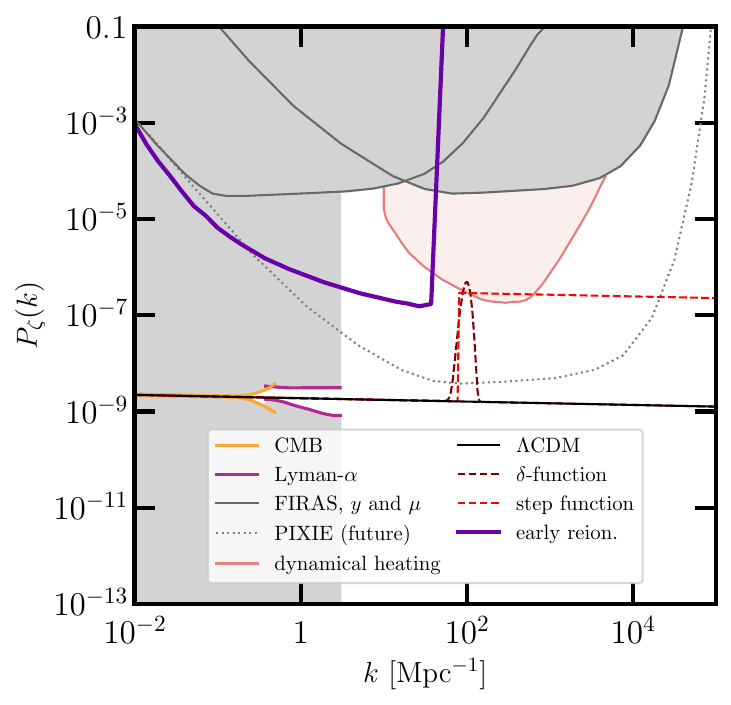}
    \caption{
        The primordial curvature power spectrum.
        We include the usual slow-roll power spectrum with parameters taken from Planck 2018~\cite{Planck:2018vyg}, labeled as ``$\Lambda$CDM", and enhancements in the form of a narrow peak and step function.
        We also show limits from CMB anisotropies~\cite{Planck:2018jri}, Lyman-$\alpha$ forest~\cite{2011MNRAS.413.1717B}, dynamical heating in ultrafaint dwarfs~\cite{Graham:2024hah}, $y$-type and $\mu$-type spectral distortions as measured by FIRAS, as well as forecasted limits from PIXIE~\cite{Cyr:2023pgw}.
    }
    \label{fig:pspecs}
\end{figure}
\begin{figure}
    \centering
    \includegraphics[width=\columnwidth]{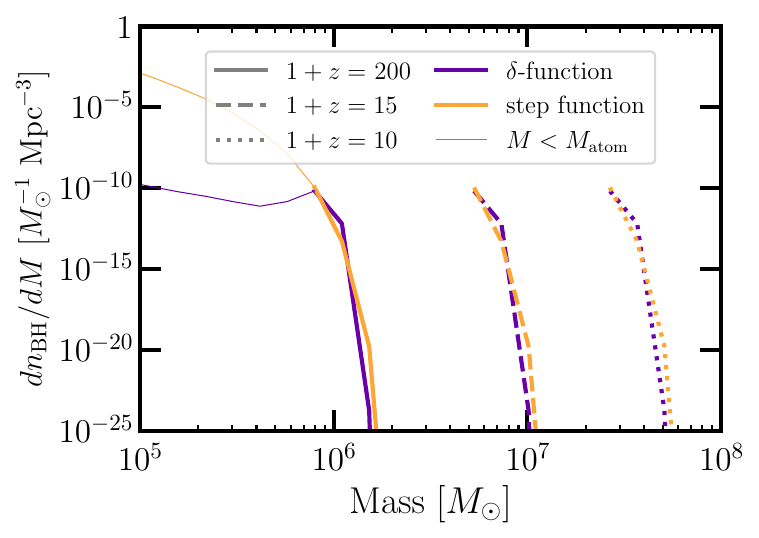}
    \caption{
        NQPBH mass function for each of the power spectra in Fig.~\ref{fig:pspecs}.
        Thin lines show halos at $1+z = 200$ below the atomic cooling mass.
        Because of accretion, the masses of the black holes grow with time and the mass function shifts towards larger masses. 
        The corresponding curves in $\Lambda$CDM are several orders of magnitude below the lowest value shown here.
    }
    \label{fig:mass_funcs}
\end{figure}

Enhancements of initial density fluctuations that are narrowly peaked at a specific scale, e.g., have a sharp peak in the primordial power spectrum, are predicted in many inflationary models that give rise to PBHs~\cite{Green:2020jor,Carr:2020xqk}.
Fig.~\ref{fig:pspecs} shows one such example. The black line shows the curvature power spectrum predicted from single-field slow-roll inflation $P_\zeta (k) = A_s (k/k_*)^{n_s - 1}$, where we take the parameters to be the best-fit values from Planck 2018: $\ln (10^{10} A_s) = 3.043$, and $n_s = 0.9652$ at the pivot scale $k_* = 0.05$ Mpc$^{-1}$. To this $\Lambda$CDM power spectrum, we add a narrow Gaussian function at $k = 100$ Mpc$^{-1}$ with a width of $\Delta k = 0.1 \times k$ that enhances the power by a factor of 310. We also show limits from CMB anisotropies~\cite{Planck:2018jri}; the Lyman-$\alpha$ forest~\cite{2011MNRAS.413.1717B}; overproduction of $y$-type and $\mu$-type CMB spectral distortions~\cite{Cyr:2023pgw}; and dynamical heating of dwarf galaxies~\cite{Graham:2024hah}.  One can also set limits from strong lensing~\cite{Gilman:2021gkj} or dwarf galaxy properties~\cite{Esteban:2023xpk,Balaji:2024wkv}, although we do not show a direct comparison here since these studies assume more specific forms for the enhancements. 
Our enhanced model lies well outside of most of the constraints, but is nearly constrained by the dynamical heating bounds and falls within the sensitivity limits of future spectral distortion probes. Hence, a future spectral distortion experiment such as PIXIE~\cite{2011JCAP...07..025K} may shed light on this possible SMBH formation mechanism.

The initial black hole mass function at $1+z = 200$ is shown in Fig.~\ref{fig:mass_funcs}. 
Thin lines show halos with mass below $M_\mathrm{atom}$ (the mass at which a halo virializes with $T = 10^4$ K and can cool atomically) which do not form NQPBHs.
Within $\Lambda$CDM, no modes have collapsed at this redshift, so the number density of halos is exponentially suppressed. 
However, after including the peaked enhancement, halos with masses less than a few times $10^6 \,M_\odot$ begin to collapse much sooner. There is a peak at $10^6 \,M_\odot$, which corresponds to the enhanced scale at $k = 100$ Mpc$^{-1}$; however, smaller masses also collapse due to the increased mass variance. We also show how the NQPBH mass function changes at later redshifts due to accretion. Integrating over the mass function, we find for our less optimistic angular momentum distribution that $n_\mathrm{BH} = 1.1 \times 10^{-6}$~Mpc$^{-3}$; for the more optimistic case, we obtain $n_\mathrm{BH} \sim 1.1 \times 10^{-5}$~Mpc$^{-3}$. 
These are comparable to the number densities of LRDs, approximately $1.3 \times 10^{-5}$~Mpc$^{-3}$, inferred from JWST~\cite{2024ApJ...963..129M,Pizzati:2024prz}.
Some studies report higher SMBH number densities; e.g., Ref.~\cite{Cammelli:2025woo} finds $n_{\rm BH}\gtrsim6\times10^{-3}\,\mathrm{Mpc}^{-3}$ at $z\gtrsim6$. 
Matching these values would require a larger enhancement by only an order unity factor in our models; however, we emphasize that SMBHs likely formed from multiple pathways, so NQPBHs need not account for all observed SMBHs.

To imitate models that produce broader enhancements in power at small scales, we also study a step-like feature.
For a step-function-like enhancement, the particular power spectrum we use is also shown in Fig.~\ref{fig:pspecs}: power in modes with $k > 80$ Mpc$^{-1}$ is enhanced by a factor of 180. The corresponding mass functions are also shown in Fig.~\ref{fig:mass_funcs}. Integrating over this distribution, we find a number density of $n_\mathrm{BH} = 1.7 \times 10^{-6}$~Mpc$^{-3}$ using our less optimistic spin distribution and $n_\mathrm{BH} = 1.7 \times 10^{-5}$~Mpc$^{-3}$ using our more optimistic assumption.

Although the power spectra we consider above grow more steeply than is allowed in the simplest one-field inflationary models~\cite{Byrnes:2018txb}, they serve to illustrate the required enhancement in a model-independent way. Moreover, there are myriad models where similar enhancements can occur, such as inflationary particle production, which can yield overdensities that are not well described by a power spectrum and are thus subject to weaker constraints~\cite{Kim:2021ida,Kumar:2025gon} (see also~\cite{Ezquiaga:2022qpw}). While canonical models require non-perturbative parameter values to achieve such large overdensities, it is not unreasonable to consider rare processes during inflation as providing a possible starting point.
Dark matter models, such as axions with large misalignment \cite{Arvanitaki:2019rax}, offer an alternative route to spikes in density perturbations at specific scales without appealing to inflation-era effects.

\textbf{Early reionization}---Outside of forming NQPBHs, atomic-cooling halos collapsing after $1+z \sim 200$ eventually begin to form H$_2$, triggering the first episode of star and galaxy formation~\cite{2015ApJ...814...18H,2024PASJ...76..850I}. 
If there is too much growth of structure in the early universe, then the ionizing radiation from early stars and galaxies will cause the universe to reionize far earlier than inferred from CMB~\cite{Planck:2018vyg} or absorption troughs in the spectra of distant quasars~\cite{2015MNRAS.447.3402B}. We can therefore set a constraint on enhancements to the power spectrum by requiring that cosmic structure does not in excessive abundance at these high redshifts. A measure of structure formation that is used in simple models of reionization and star formation is the \textit{collapse fraction}
\begin{equation}
    f_\mathrm{coll} (z, M_\mathrm{min}) = \frac{1}{\bar{\rho}} \int_{M_\mathrm{min}}^\infty \D M \, M \frac{\D n}{\D M} (z),
\end{equation}
where $\bar{\rho}$ is the mean matter density.
$\Lambda$CDM predicts that the collapse fraction at $1+z = 20$ should be $\lesssim 10^{-3}$.

In Ref.~\cite{Furlanetto:2004nh}, the ionized fraction of hydrogen follows
\begin{equation}
    x_\mathrm{HII} \equiv \frac{n_\mathrm{HII}}{n_\mathrm{H}} = \zeta f_\mathrm{coll} (z, M_\mathrm{atom}),
\end{equation}
where $n_\mathrm{HII}$ is the number density of ionized hydrogen and $\zeta$ is an ionization efficiency factor. The universe is completely reionized when $f_\mathrm{coll} (z_\mathrm{reion}, M_\mathrm{atom}) = \zeta^{-1}$. We will assume $\zeta = 10$, which is among the lowest values quoted in the literature~\cite{Furlanetto:2004nh,2009ApJ...703L.167A,Binnie:2025mrf}.
Since this is a relatively simple model of reionization, to ensure that our limits are not artificially strong, we only rule out very early endings to reionization, e.g. before $1+z = 20$. 
Then our condition that the universe not be reionized too early is
\begin{equation}
    f_\mathrm{coll} (1+z = 20, M_\mathrm{atom}) < 0.1 .
    \label{eqn:reion_cond}
\end{equation}
where $M_\mathrm{atom} = 2 \times 10^7 \,M_\odot$ at this redshift.

We constrain the power spectrum at each wavenumber using narrow peak enhancements and find the amplitude at which Eq.~\eqref{eqn:reion_cond} is saturated---our procedure is described in more detail in the end matter.
The resulting limit is shown in Fig.~\ref{fig:pspecs} as the solid purple line. 
The limits we set here are stronger than existing power spectrum constraints from FIRAS data and stronger than PIXIE forecasts for $k \lesssim 0.1$ Mpc$^{-1}$~\cite{Cyr:2023pgw}. Note that our fiducial models are not excluded by this condition.

\textbf{Conclusion}---We have described a novel mechanism for forming SMBH seeds at $z \sim 200$. Dark matter halos forming deep in the cosmic dark ages at $1+z \gtrsim 200$ may meet the conditions for direct collapse, and such halos can form from enhanced density fluctuations that are similar to those needed to form PBHs but much smaller in amplitude. We demonstrate using a power spectrum enhanced with a narrow peak and a step function that these NQPBHs can explain the number density of high-$z$ black holes inferred from JWST while avoiding existing constraints, namely those from CMB spectral distortions. Hence, NQPBHs provide a new target for future instruments that observe CMB spectral distortions or allow access to extremely high $z$. 
We also show a constraint on the primordial power spectrum derived by requiring that the early atomic-cooling halos associated with these enhanced modes not reionize the universe too early.

A compelling feature of the mechanism proposed here is that it does not require the extreme density enhancements and fine-tuning invoked in many models that produce PBHs~\cite{Cole:2023wyx}. 
However, while NQPBHs avoid some of the problems facing PBHs and can in principle account for the observed number density of SMBHs at late times, they also share some of the same issues, such as how they later migrate to the centers of galaxies~\cite{Ma:2021ivz,2022MNRAS.510..531C,Ziparo:2024nwh}. 

Given the uncertainties surrounding how compact objects form from direct collapse at $1+z \gtrsim 100$, there are several immediate directions for the follow-up studies. 
For example, high-redshift cosmological simulations including hydrodynamics and feedback in these large overdensities will
determine if the halos avoid fragmentation even in the late stages of NQPBH formation, as well as determine if the halos can sustain the high gas infall rates needed to form heavy black hole seeds. 
Moreover, since it is the environments of SMBHs, e.g. their host galaxies, that encode information about their formation, predictions from simulations of the evolution of these objects is key for exploring observational signatures of NQPBHs.  
In addition, although we employ the Press-Schechter formalism, it is known to underestimate the abundance of the largest halos compared to simulations; hence it would be instructive to explore other mass functions~\cite{Sheth:1999mn,2001MNRAS.323....1S}. 

The ubiquitousness of SMBHs detected by JWST before $z \sim 4-7$, along with the fact that their abundances vastly exceed theoretical predictions, is highly suggestive of the existence of multiple pathways to produce initial BH seeds. Confirmation of NQPBHs would not only shed light on the origin of SMBHs but also have profound implications regarding the initial conditions of the universe.

\textit{Acknowledgements}---We thank David Curtin, M. Sten Delos, Zoltan Haiman, Julien Lavalle, Vivian Poulin, Harikrishnan Ramani, Ravi Sheth, Magdalena Siwek, and Ken Van Tilburg for useful discussions regarding this work.
W.Q. is supported by the Simons Society of Fellows through Grant No. SFI-MPS-SFJ-00006250. S.K. and N.W. are supported in part by the National Science Foundation grant PHY-2210498 and the Simons Foundation. N.W. is supported by the US-Israel Binational Science Foundation (BSF) under grant 2022287. P.N. acknowledges support from the Gordon and Betty Moore Foundation (Grant \#8273.01) and the John Templeton Foundation (Grant \#62286) that fund the Black Hole Initiative (BHI) at Harvard University. P.N. also acknowledges support from the John Templeton Foundation via Grant \#126613 for this work.

\onecolumngrid
\begin{center}
    \textbf{End Matter}
\end{center}
\twocolumngrid

\begin{figure}
    \centering
    \includegraphics[width=\columnwidth]{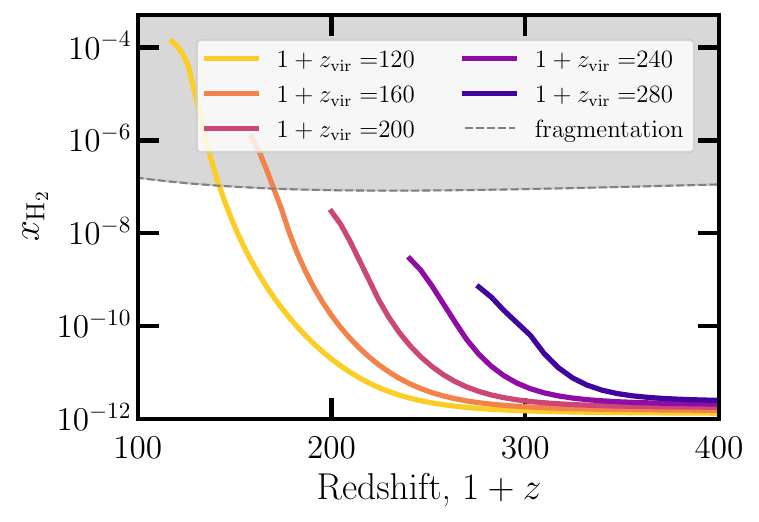}
    \caption{
        The molecular hydrogen abundance in halos virializing at different redshifts.
        The gray contour shows the values of $x_{\mathrm{H}_2}$ for which the halo is likely to fragment and form stars instead of NQPBHs.
        We find that halos virializing at $1+z_\mathrm{vir} > 200$ are therefore capable of forming NQPBHs.
    }
    \label{fig:H2_est}
\end{figure}

\textit{Limits on the formation redshift for NQPBHs}---To determine the critical amount of H$_2$ that must form in a dark matter halo before its baryonic content (gas) begins to significantly fragment, and therefore prevent direct collapse, one can compare free-fall time of the halo, $\tau_{ff} = \sqrt{3\pi / (32 G \rho)}$, to the molecular hydrogen cooling time, $\tau_{\mathrm{H}_2} = T / \dot{T}_{\mathrm{H}_2}$~\cite{Wolcott-Green:2011tul}. The molecular cooling rate is given by $\dot{T}_{\mathrm{H}_2} = n_{\mathrm{H}_2} \Lambda_{\mathrm{H}_2}$, where we use the cooling function $\Lambda_{\mathrm{H}_2}$ from Ref.~\cite{Galli:1998dh}. We calculate these timescales by assuming a dark matter halo temperature of $T = 10^4$~K, the threshold for atomic cooling, and a halo density corresponding to the virial density, which is $18 \pi^2$ times the mean density of the universe at that time. Fig.~\ref{fig:H2_est} shows the region where $\tau_{ff} > \tau_{\mathrm{H}_2}$ across $100 < 1+z < 400$ marked as a grey contour. We find within this redshift range that the critical molecular hydrogen abundance is $x_{\mathrm{H}_2} \gtrsim 10^{-7}$. This is smaller than the value of $x_{\mathrm{H}_2} \gtrsim 10^{-3}$ quoted for star-formation in previous studies~\cite{Tegmark:1996yt, Abel:2001pr, Glover:2008pz} because we are studying halos with higher virial temperatures and $\Lambda_{\mathrm{H}_2}$ increases as a function of $T$.

We then evolve the H$_2$ abundance in these dark matter halos using the spherical top-hat approximation as described in Ref.~\cite{Qin:2023kkk}, which is closely based on the procedure outlined in Ref.~\cite{Tegmark:1996yt}. We illustrate the fate of these halo by choosing a few different values for the redshift at which the halo virializes, $z_\mathrm{vir}$, and also fix the halo mass such that $T_\mathrm{vir} = 10^4$~K according to Eq.~\ref{eqn:Tvir}. The evolution of each halo is truncated shortly after virialization, since this method is not equipped to evolve the halo deep into nonlinear collapse. In principle, one should evolve the halo for longer as above $T_\mathrm{vir} = 10^4$~K and $n_\mathrm{H} = 10^4$~cm$^{-3}$, the halo reaches a ``zone of no return" in which the H$_2$ abundance remains suppressed from collisional dissociation~\cite{Inayoshi:2012zi,Fernandez:2014wia}.
The H$_2$ abundance will continue to rise if we evolve the halos to this higher density; however, we anticipate that this will only narrow the the redshift range in which it is possible to form NQBPHs without closing it entirely. 
Fig.~\ref{fig:H2_est} shows the $x_{\mathrm{H}_2}$ in each of these halos as a function of redshift. As halos virialize later, the abundance of $x_{\mathrm{H}_2}$ increases because the CMB cools to the point where it can no longer sufficiently dissociate the intermediate H$^-$ state~\cite{Hirata:2006bt}.
We therefore find that halos virializing prior to $1+z_\mathrm{vir} \sim 200$ are unlikely to fragment and can potentially host NQPBHs.

One might be concerned that since these halos are unusually dense, they may be opaque to the CMB and the H$_2$ forming within may be shielded from dissociation.
We have explicitly checked that the diffusion times ($\tau_\mathrm{diff} = R_\mathrm{halo}^2 n_e \sigma_T / c$) for CMB photons in the halos considered in Fig.~\ref{fig:H2_est} are orders of magnitude smaller than the halo free-fall times---hence the CMB photons permeate the halo and may destroy H$_2$.
In addition, the $\tau_{\mathrm{H}_2}$ comparison described above does not account for the fact that dynamical heating may also help stabilize the halo against fragmentation~\cite{Yoshida:2003rw,2019Natur.566...85W}; given that including this effect \textit{favors} the formation of NQPBHs, we neglect it here to ensure that our conditions for NQPBH formation are sufficiently stringent and hence conservative.

There is a limited range of past literature that has also studied the redshift threshold before which molecular cooling is suppressed; however, these studies have been primarily interested in the star-forming properties of enhanced density perturbations, as opposed to the consequences of black hole formation.
For example, while we use a toy halo model to estimate the critical redshift, Ref.~\cite{2015ApJ...814...18H} uses cosmological hydrodynamic simulations and is in agreement that atomic-cooling halos form at $1+z \gtrsim 200$, 
More recently, Ref.~\cite{2024PASJ...76..850I} extended the study of these halos to much higher gas densities using a one-zone model and found a somewhat higher redshift of $1+z \gtrsim 500$.
We leave the determination of this precise window as a direction for future study.

%


\textit{Relative motion of baryons relative to dark matter in the early universe}---While the timing of the dark matter halo formation is determined by the initial value of the over-density, this does not mean that baryons in all halos will be able to immediately fall into the dark matter gravitational potential.
Ref.~\cite{2010PhRvD..82h3520T} showed that after decoupling, the baryons will have a velocity offset relative to the dark matter that varies spatially, and in places where this relative velocity (sometimes called the ``streaming velocity") is large, it will be more difficult for baryons to fall into the dark matter potentials and this is expected to therefore delay structure formation.

The distribution of relative velocities is well-approximated by a three-dimensional Gaussian distribution, and the root-mean-square streaming velocity of this distribution at recombination is about $v_{s,i} = 30$~km/s~\cite{2010PhRvD..82h3520T}. 
In addition, the streaming velocities decay with redshift as $v_s(z) = v_{s,i} \times (1+z) / 1100$.
For comparison, the typical velocity of gas at a temperature of $10^4$ K is given by the sound speed, $c_s = \sqrt{\frac{5 T}{3 \mu m_p}} \approx 11$ km/s. 
By redshift of $z \sim 400$, the typical streaming velocity is equal to this value and, by redshift of $z \sim 200$, the typical velocity will have dropped to $v_s(z) \sim 5.5$ km/s.

We emphasize that the streaming velocities follow a distribution, so even if a density fluctuation virializes before $1+z \sim 400$, the baryons in the halo may still be able to fall into the gravitational potential. 
For example, at $z \sim 1100$, about $1\%$ of halos will have a streaming velocity that is less than the sound speed.
By $z \sim 500$, this fraction has already grown to $10\%$, and by $z \sim 200$ it is $71\%$.
Hence, the formation of the NQPBHs is not prohibited at higher redshifts, but care must be taken in calculating their rate of formation.

\textit{Accretion history}---Assuming the NQPBH forms with an order unity fraction of the halo's mass, we can evolve the black hole's mass, as well as map between mass and luminosity, by assuming an accretion history. While the accretion history of black holes at such high redshifts is highly uncertain, we will make headway by assuming the black hole accretes optimally at the Eddington limit. The largest luminosity that a black hole can have before radiation pressure overcomes gravity is given by
\begin{equation}
    L_\mathrm{Edd} = \frac{4\pi G M m_p c}{\sigma_T} .
    \label{eqn:L_edd}
\end{equation}
Since the accretion rate of the black hole is related to its luminosity by $L = \epsilon \dot{M}_\mathrm{BH} c^2$, where $\epsilon$ is the efficiency with which the accreted mass is converted into radiation, then the mass of the black hole grows exponentially as
\begin{equation}
    M (t) = M_i \exp \left( \frac{4\pi G m_p c}{\epsilon \sigma_T} f_\mathrm{duty} t \right) ,
    \label{eqn:Medd_over_time}
\end{equation}
where we have also introduced the duty cycle parameter $f_\mathrm{duty}$, which characterizes the amount of the time that the black hole is actively accreting.
We will assume $f_\mathrm{duty} = 0.3$, as it is unlikely that these black holes will be steadily accreting at the Eddington limit as far back as the Dark Ages.
Hence, $M_\mathrm{JWST}$ is the black hole mass that would have an Eddington luminosity equal to the JWST luminosity threshold of $L > 10^{45}$~erg~s$^{-1}$~\cite{Pizzati:2024prz}, and the minimum mass in Eqn.~\eqref{eqn:nBH} is therefore given by $M_\mathrm{min} = \max [M_\mathrm{JWST}, M_\mathrm{atom}]$.

The resulting number density of NQPBHs will change if the true accretion history deviates from these assumptions. 
For example, if these black holes experience enhanced accretion than assumed here, due to e.g. periods of super-Eddington accretion, then the black holes must initially be very small, which requires SMBH seeds to form in the parameter space on the right side of Fig.~\ref{fig:summary} and thus requires more extreme over-density enhancements to form NQPBHs.
If these black holes experience less accretion due to e.g. lower duty cycles, then the seed black holes can be heavier.
In addition, our results may change with a more realistic accretion model that accounts for radiative feedback, such as Park-Ricotti~\cite{2013ApJ...767..163P}---whether the net result is to increase or decrease the total mass accreted depends on the velocity of the black holes, as well as the local sound speed (see e.g. Ref.~\cite{Agius:2024ecw}, which shows how different accretion models affect constraints on PBHs).
We leave a detailed study of the impact of accretion modeling on NQPBHs to future work.

\textit{Spin parameter distribution of early dark matter halos}---If we assume that the spin distribution of dark matter halos is independent of their mass, then Eqn.~\eqref{eqn:nBH} factorizes into
\begin{gather}
    n_\mathrm{BH} = \mathcal{N}_\lambda \int_{M_\mathrm{min}}^\infty dM \, \kappa \frac{dn}{dM} \bigg \rvert_{1+z = 200} , \\
    \mathcal{N}_\lambda = \int_0^{\lambda_\mathrm{max}} d\lambda \, P(\lambda), 
    \label{eqn:nBH_sep}
\end{gather}
where $P(\lambda)$ is the probability distribution of the spin parameter.
We can thus separate out the integral over spin, which becomes an additional normalization factor for the halo mass function.

While $P(\lambda)$ is known to be well-approximated by a log-normal function, with $\langle \lambda \rangle \sim 0.04$ and $\sigma \sim 0.55$ at $1+z = 10$~\cite{1992ApJ...399..405W,Mo:1997vb,Bullock:2000ry,Jang-Condell:2000wmt,vandenBosch:2001bp,Davis:2008nn}, to our knowledge there are no studies of this distribution at redshifts above $1+z \sim 100$.
However, given that the angular momentum of these systems grows with time due to tidal torques~\cite{Hoyle1949,1969ApJ...155..393P,1970Afz.....6..581D,1979MNRAS.186..133E,1984ApJ...286...38W,2002MNRAS.332..325P,2002MNRAS.332..339P}, then at the high redshifts we are considering more halos will meet the angular momentum requirements to form NQPBHs than at cosmic dawn.
To show the possible range of impact of the spin parameter distribution, we will take the spin-distribution at $1+z = 200$ to be either the same as at $1+z = 10$, which likely underestimates the abundance of black holes, or use a lower angular momentum log-normal distribution with $\langle \lambda \rangle = 0.01$ and $\sigma = 0.25$. While the exact angular momentum threshold for black hole formation is also uncertain, we will use the halos from Ref.~\cite{2006MNRAS.371.1813L} as a guideline and take $\lambda_\mathrm{max} = 0.02$ as our angular momentum limit.
In this case, the former spin parmeter distribution gives $\mathcal{N}_\lambda = 0.10$ and the latter gives $\mathcal{N}_\lambda = 0.999$.

As a final remark, with the additional power at small scales, the spin parameter distribution may also have a scale-dependent feature with higher amplitude. 
If this causes the overdensities considered here to form a larger disk than in a typical $\Lambda$CDM halo, then the extended self-gravitating disk would be more likely to fragment.
We leave a detailed investigation of this as a direction for future consideration as this will need to addressed with a cosmological large volume simulation to capture tidal fields and their effects accurately.

\textit{Shooting method}---Using Eqn.~\eqref{eqn:reion_cond}, we can constrain the primordial power spectrum as follows.
At each $k$, we construct a power spectrum with a narrow peak at that wavenumber, as described in the text, and calculate $f_\mathrm{coll} (1+z = 20, M_\mathrm{atom})$ for no enhancement and a factor of $10^9$ enhancement.
We recalculate $f_\mathrm{coll} (1+z = 20, M_\mathrm{atom})$ at the geometric mean of the two amplitudes and continue until we find the peak amplitude that satisfies $f_\mathrm{coll} (1+z = 20, M_\mathrm{atom}) = 0.1$ to within a few percent.

\bibliography{references}
\end{document}